\documentclass[12pt,fleqn]{article}

\usepackage{amsmath,amssymb,graphicx}

\usepackage[textwidth=15.9cm,textheight=23.2cm]{geometry}

\allowdisplaybreaks[1]

\numberwithin{equation}{section}

\newcommand\jhep[3]{{\it J. High Energy Phys.\ }{\bf #1} (#2) #3}

\newcommand\npb[3]{{\it Nucl.\ Phys.\ }{\bf B#1} (#2) #3}

\newcommand\nc[3]{{\it Nuovo Cim.\ }{\bf #1} (#2) #3}

\newcommand\prd[3]{{\it Phys.\ Rev.\ }{\bf D#1} (#2) #3}

\newcommand\prl[3]{{\it Phys.\ Rev.\ Lett.\ }{\bf #1} (#2) #3}

\newcommand{\hepth}[1]{{\tt hep-th/#1}}

\begin{document}


\thispagestyle{empty}
\renewcommand{\thefootnote}{\fnsymbol{footnote}}

\bigskip\bigskip

\begin{center} \noindent \Large \bf
Less is More:\\ Non-renormalization Theorems from \\ Lower
Dimensional Superspace
\end{center}

\bigskip\bigskip\bigskip

\bigskip\bigskip\bigskip

\centerline{ \normalsize \bf Z. Guralnik$^{\,a}$, S. Kovacs$^{\,b}$
and B. Kulik$^{\,b}$~\footnote[1]{\noindent \tt
zack@physik.hu-berlin.de, bogdan.kulik@aei.mpg.de,
stefano.kovacs@aei.mpg.de} }

\bigskip
\bigskip\bigskip

\centerline{$^a$ \it Institut f\"ur Physik} \centerline{\it
Humboldt-Universit\"at zu Berlin} \centerline{\it Newtonstra{\ss}e
15} \centerline{\it 12489 Berlin, Germany}
\bigskip
\centerline{$^b$ \it Max-Planck-Institut f\"ur Gravitationsphysik}
\centerline{\it Albert-Einstein-Institut} \centerline{\it Am
M\"uhlenberg 1, D-14476 Golm, Germany}
\bigskip\bigskip

\renewcommand{\thefootnote}{\arabic{footnote}}

\centerline{\bf \small Abstract}
\medskip

\noindent
{\small We discuss a new class of non-renormalization theorems in
${\cal N}=4$ and ${\cal N}=2$ Super-Yang-Mills theory, obtained by
using a superspace which makes a lower dimensional subgroup of the
full supersymmetry manifest.  Certain Wilson loops (and Wilson
lines) belong to the chiral ring of the lower dimensional
supersymmetry algebra, and their expectation values can be
computed exactly.}

\newpage

\section{Introduction}

In four dimensional theories with extended supersymmetry,  it is
simple to make an ${\cal N}=1$ subgroup of the supersymmetries
manifest and possible, albeit more complicated, to make ${\cal
N}=2$ manifest using harmonic superspace.  The advantages of an
off-shell formulation which make as much supersymmetry as possible
manifest include demonstrations of non-renormalization theorems as
well as calculational efficiency.

In the following,  we will make supersymmetries other than ${\cal
N}=1$ manifest,  but not by using a harmonic superspace. Instead,
we will make use of the fact that four-dimensional extended
supersymmetry algebras have lower-dimensional subalgebras with
four supercharges and an associated superspace which is a
dimensional reduction of the familiar four-dimensional ${\cal
N}=1$ superspace. By a lower dimensional subalgebra,  we mean a
subalgebra which does not include momenta in all four directions.
Despite the simplicity of the lower dimensional superspace, a
useful part of the full supersymmetry which does not belong to
four dimensional ${\cal N}=1$ is realized off-shell.

One of the general features of such lower dimensional superspaces
is that certain kinetic terms (in directions transverse to the
superspace) appear in a superpotential rather than a K\"ahler
potential.  Furthermore gauge connections in these directions are
bottom components of chiral superfields.  As usual, supersymmetry
leads to strong constraints on the holomorphic sector involving
chiral superfields, which does not include gauge connections and
kinetic terms in the usual four-dimensional ${\cal N}=1$
formalism.

In the case of ${\cal N}=4$ super Yang-Mills theory,  the lower
dimensional superspace formalism which we will use makes certain
supersymmetries manifest which are not accessible to either ${\cal
N}=1$ superspace or even ${\cal N} =2$ harmonic superspace.
Specifically, we will express the  ${\cal N}=4$ theory in terms of
${\cal N}=4$, $d=1$ superspace.  This will allow us to prove  and
extend a non-renormalization theorem for BPS Wilson loops which
was conjectured in \cite{Zarembo:2002an}.  A more detailed
discussion appeared in \cite{Guralnik:2003di}.

In the case of four dimensional ${\cal N}=2$ super Yang-Mills
theories, we will use ${\cal N}=2$, $d=3$ superspace to obtain exact
results for expectation values of straight BPS Wilson lines with
scalar components of hypermultiplets at the endpoints. These
expectation values are non-trivial on the Higgs branch,  and can
be expressed exactly in terms of expectation values of local
operators parameterizing the Higgs branch \cite{Guralnik:2004ve}.

\section{Warming up:  A free ${\cal N}=2$, $d=4$ hypermultiplet in
${\cal N}=2$, $d=3$ superspace}

The first example of supersymmetric action written using lower
dimensional superspace appeared in \cite{Marcus:1983wb}.  Since
then, this formalism has been applied to many supersymmetric
theories in various different contexts
\cite{Arkani-Hamed:2001tb,Erdmenger:2002ex,Erdmenger:2003kn,
Constable:2002vt,Constable:2002xt,Dijkgraaf:2003xk,Bena:2003tf}.
We now describe a very simple example,  in which a free ${\cal
N}=2$ hypermultiplet in four dimensions is described by an action
in ${\cal N}=2$, $d=3$ superspace.

In the absence of central extensions, the ${\cal N}=2$, $d=4$
supersymmetry algebra is
\begin{equation}
\{Q_{i\alpha},\bar Q^j{}_{\dot\beta} \} =
2\sigma^\mu_{\alpha\dot\beta} P_\mu \delta_i^j \, ,
\qquad \{Q_{i\alpha}, Q_{j\beta} \} = \{\bar Q^i{}_{\dot\alpha},
Q^j_{\dot\beta} \} =0 \,,
\end{equation}
where $i=1,2$. Defining
\begin{align}
{\rm Q}_{\alpha} \equiv \frac{1}{2}( Q_{1\alpha} + \bar
Q^1{}_{\dot\alpha})+ i\frac{1}{2}( Q_{2\alpha} + \bar
Q^2{}_{\dot\alpha}) \, ,
\end{align}
one finds an ${\cal N}=2$, $d=3$ subalgebra,
\begin{align}
&\{{\rm Q}_\alpha, \bar {\rm Q}_\beta \} =
2\sigma^M_{\alpha\beta}P_M \qquad M=0,1,3 \nonumber \\
&\{{\rm Q}_\alpha, {\rm Q}_\beta \} = \{ \bar {\rm Q}_\alpha, \bar
{\rm Q}_{\beta} \} =0 \, .
\end{align} A four-dimensional ${\cal N}=2$ theory can be written
in terms of the associated ${\cal N}=2$, $d=3$ superspace. This
superspace is equivalent to the dimensional reduction of the
familiar ${\cal N}=1$, $d=4$, and is spanned by $x^0,x^1,x^3,
\theta, \bar\theta$. The ${\cal N}=2$, $d=3$ superfields necessary
to describe a ${\cal N}=2$, $d=4$ theory have the general form
$F(x^0,x^1,x^3, \theta,\bar\theta| x^2)$, where the spatial
coordinate $x^2$ transverse to the superspace should be thought of
as a continuous index labeling an infinite number of ${\cal N}=2$,
$d=3$ superfields.

A free massless ${\cal N}=2$, $d=4$ hypermultiplet can be built from
two ${\cal N}=2$, $d=3$ chiral multiplets $\Phi_1(x^2)$ and
$\Phi_2(x^2)$, which are annihilated by the superspace derivative
\begin{align}
\bar D_{\alpha} \equiv -\frac{\partial}{\partial \bar
\theta^\alpha} - i
\theta^{\beta}\sigma^{\mu}_{\beta\alpha}\partial_{\mu},\, \, \mu
=0,1,3 \, .
\end{align}
The action is
\begin{align}
S =\int dx^2 \int dx^0 dx^1 dx^3 \left[\int d^4\theta\,
(\bar \Phi_1 \Phi_1 + \bar \Phi_2 \Phi_2)
+\int d^2\theta\, \Phi_1 \frac{\partial}{\partial X^2}
\Phi_2 \, +\, {\rm c.c.} \right] \, .
\end{align}
Note that the K\"ahler potential by itself only gives rise to
kinetic terms in the $0,1,3$ directions belonging to the
superspace. Remarkably, the kinetic terms in the $x^2$ direction
arise from the superpotential rather than the K\"ahler potential.
Herein lies the advantage of a lower dimensional superspace over
the familiar ${\cal N}=1$, $d=4$ superspace:  one can exchange
K\"ahler terms for superpotential terms.  This is a strong hint
that the lower-dimensional superspace formalism can be used to
find new non-renormalization theorems. We will now discuss two
examples of non-renormalization theorems which have been found
using the lower-dimensional superspace formalism.

\section{Chiral Wilson loops in ${\cal N}=4$ SYM}

The four-dimensional ${\cal N} =4$ supersymmetry algebra contains
a one dimensional ${\cal N}=4$ subalgebra containing four real
supercharges. The superspace associated with this subalgebra is
equivalent to a dimensional reduction of familiar four dimensional
${\cal N}=1$ superspace, and is spanned by $t, \theta,
\bar\theta$.  When writing the four-dimensional ${\cal N}=4$
theory in this superspace,  a generic superfield has the form
$F(t, \theta,\bar\theta | \vec x)$,  where the spatial coordinates
$\vec x = x^{1,2,3}$ are continuous indices labeling an infinite
number of ${\cal N}=4$, $d=1$ superfields.  The superfield content
of the theory consists of chiral superfields $\Phi_i(\vec x)$ for
$i=1,2,3$ and vector superfields $V(\vec x)$. Although this is
similar to the superfield content in the more familiar ${\cal
N}=1$, $d=4$ superspace,  the way component fields are distributed
among the ${\cal N}= 4$, $d=1$ superfields is very different, as we
will see shortly. The action in ${\cal N}=4$, $d=1$ superspace is
\begin{align}
\label{other}
S = & \frac{1}{g^2} \int d^3x\, \int dt \left\{ \int d^2\theta\,
{\rm tr}\left[ {\cal W}_{\alpha}{\cal W}^{\alpha} + \epsilon_{ijk}
(\Phi_i\frac{\partial}{\partial x^j} \Phi_k +
\frac{2}{3}\Phi_i\Phi_j\Phi_k) \right] + {\rm c.c.} \right.
\nonumber\\
& \hspace*{2.5cm} \left. + \int d^4\theta\,{\rm tr}\,\bar\Omega_i
e^V\Omega_i e^{-V} \right\} \, ,
\end{align}
where
\begin{align} \Omega_i \equiv \Phi_i + e^{-V}(i\partial_i -
\bar\Phi_i)e^{V}\, .
\end{align}
Note that the index $i=1,2,3$ labeling chiral superfields has been
``identified'' with the spatial index (similar to the identification
of spatial and lie algebra indices in the 't Hooft Polyakov
monopole). The bottom components of the chiral superfields are
\begin{align}
\Phi_i|_{\theta=\bar\theta=0} = A_i + i X^i \, ,
\end{align}
where $A_i$ are gauge connections in the spatial directions
transverse to the superspace,  and $X^i$ are three (out of the
six) adjoint hermitian scalars. The rest of the adjoint scalars
are contained in the vector superfield.  Under gauge
transformations parameterized by chiral superfields $\Lambda(\vec
x)$,
\begin{align} e^{V} \rightarrow
e^{i\Lambda^{\dagger}}e^{V}e^{-i\Lambda},\,\,\,\,\,\,\, \Phi_i
\rightarrow e^{i\Lambda} \Phi_i e^{-i\Lambda} - e^{i\Lambda}
i\frac{\partial}{\partial x^i}e^{-i\Lambda} \, .
\end{align}

A special class of Wilson loops are chiral superfields in ${\cal
N}=4$, $d=1$ superspace,
\begin{align}
{\cal W}(C) \equiv {\rm tr} {\cal P}\left( e^{i\oint_C \Phi_i
dx^i}\right) \, ,
\end{align}
which have bottom components
\begin{align}
\label{loops}
{\cal W}(C)|_{\theta=\bar\theta =0} = {\rm tr} {\cal P}\left(
e^{i\oint_C (A_i + i X^i) dx^i}\right)\, .
\end{align}
The latter belong to a class of BPS Wilson loops in the ${\cal N}
=4$ theory originally discussed in \cite{Zarembo:2002an}.  These
in turn belong to a class of ``locally BPS'' Wilson loops
containing adjoint scalars which were introduced in the context of
AdS/CFT duality
\cite{Maldacena,Rey:1998ik,Berenstein:1998ij,Drukker:1999zq}.
The loops (\ref{loops}) preserve $1/4$ or $1/8$ of the $16$
supersymmetries of the theory, depending on whether the path $C$
lies in a two or three-dimensional subspace.  In
\cite{Zarembo:2002an},  it was conjectured that the expectation
values of $1/4$ BPS loops are un-renormalized in the large $N$
limit.  This conjecture was based on perturbative calculations as
well as strong coupling results obtained using the AdS/CFT
duality. In fact one can show that both the $1/4$ and $1/8$ BPS
loops are not renormalized, even at finite $N$,  by using the fact
that they belong to a chiral ring with respect to ${\cal N} =4$,
$d=1$ supersymmetry.  The chiral ring is constrained by the
superpotential,  which in this case resembles a Chern-Simons
action and has a three-dimensional diffeomorphism invariance.

Consider a variation
\begin{align}\label{trans}
\Phi_i(\vec x) \rightarrow \Phi_i(\vec x)
+ \epsilon g_{i, \vec x}[\Phi]\, ,
\end{align}
where the small parameter $\epsilon$ is a chiral superfield and
$g_{I,\vec x}$ is an arbitrary functional of chiral superfields. The
equations of motion, $\delta S=0$, which follow from this variation are
\begin{align}\label{Kon}
\bar D^2 {\rm tr} \left(g_{i,\vec x}[\Phi](e^{-V(\vec
x)}\bar\Omega_i(\vec x) e^{V(\vec x)} - \Omega_i(\vec x) ) \right)
+ {\rm tr} g_{i,\vec x}[\Phi]\frac{\delta W}{\delta \Phi_i(\vec
x)} = {\cal A}\, ,
\end{align}
where ${\cal A}$ is a possible anomalous term which vanishes
classically. We will choose $g_{i,\vec x} [\Phi]$ to be a spatial
Wilson line on a contour ${\cal C}_{\vec x}$ which begins and ends at
the point $\vec x$,
\begin{align}
g_{i,\vec x}[\Phi] = W(C,\vec x) = P \exp\left( i\oint_{{\cal
C}_{\vec x}} \vec \Phi\cdot d\vec y\right)\, .
\end{align}
Equation (\ref{Kon}) then becomes
\begin{align}\label{Konam}
\bar D^2( \cdots ) = {\rm tr} \left( W(C,\vec x) \epsilon_{ijk}
{\cal F}_{jk}(\vec x)\right) + {\cal A}\, ,
\end{align}
where \begin{align} {\cal F}_{jk} = \partial_j\Phi_k -
\partial_k\Phi_j + i[\Phi_j,\Phi_k]\, . \end{align}
The relation $(\bar D^2 J)|_{\theta=\bar\theta =0} = [\bar Q,
[\bar Q, J|_{\theta=\bar\theta=0}]$ implies that  $\langle\bar D^2
J|_{\theta=\bar\theta=0}\rangle =0$ in a supersymmetric vacuum.
Therefore (\ref{Konam}) implies
\begin{align}\label{expect}
\left< \left. {\rm tr} \left( W(C,\vec x) \epsilon_{ijk} {\cal
F}_{jk}(\vec x)\right) \right|_{\theta = \bar\theta =0} \right> =
\langle{\cal A}|_{\theta = \bar\theta =0}\rangle \, .
\end{align}
It can be shown that the anomalous term vanishes
\cite{Guralnik:2003di}. This may seem surprising to the reader
expecting a non-trivial Konishi anomaly \cite{Konishi:1985tu}. The
absence of an anomaly is related to the fact that the fermionic
components of the ${\cal N}=4$, $d=1$ chiral superfields $\Phi_i(\vec
x)$ are not of definite chirality from the point of view of the
four-dimensional Lorentz group, unlike the fermionic components of
${\cal N}=1$, $d=4$ superfields. For ${\cal A} =0$, (\ref{expect})
implies shape invariance\footnote{This is actually a larger
symmetry than diffeomorphism invariance, which does not relate
knots of different topology. Breaking shape invariance to
diffeomorphism invariance would require a non-zero anomaly.} of
the Wilson loop expectation value, since the insertion of the
field strength ${\cal F}_{jk}$ generates an infinitesimal
deformation of the contour ${\cal C}$ in the $jk$ plane.

Shape invariance of the Wilson loop expectation value leads to the
conclusion
\begin{align}\label{result}
\left<\frac{1}{N}{\rm tr} P \exp\left(i\oint\, \phi\right)\right>
= 1\, .
\end{align}
Note that because of conformal invariance,  this can not be shown
simply by shrinking the Wilson loop.  Instead we use the following
argument. Given a Wilson-loop $\frac{1}{N}{\rm tr} W$ associated
with a path $C$ in $\mathbb{R}^3$ one can smoothly deform the path
within $\mathbb{R}^3$ such that it goes around $C$ multiple times
and the Wilson loop becomes $\frac{1}{N}{\rm tr} W^n$ for any
$n>1$\footnote{Deforming to n=0 can not be done in a smooth way
without introducing a cusp.}. Shape independence implies that the
expectation value is unchanged,
\begin{align}\label{shp}
\langle\frac{1}{N}{\rm tr} W^n\rangle=\langle\frac{1}{N}{\rm tr}
W\rangle\, .  \end{align} Furthermore, there are relations amongst
the variables ${\rm tr} W^n$ of the form
\begin{align}\label{relns}
{\rm tr} W^n(C) = \sum {\rm tr}W^q (C){\rm tr}W^p(C) \cdots\,
,\end{align}  such that ${\rm tr} W^n$ for $n=1,\ldots,2N^2$ form
a complete independent set. This follows from the fact that $W$ is
an $N\times N$ matrix with complex entries and no constraints.
Note that, unlike the usual Wilson loop,  the chiral Wilson loop
is not the trace of a unitary matrix because the exponent involves
both hermitian and anti-hermitian parts. Since these Wilson loops
belong to a chiral ring, the expectation values factorize,
\begin{align}\label{factor}
\langle\frac{1}{N}{\rm tr} W^n \frac{1}{N}{\rm tr} W^m\rangle =
\langle\frac{1}{N}{\rm tr} W^n\rangle\langle \frac{1}{N}{\rm tr}
W^m\rangle \, .
\end{align}
The relations (\ref{relns}) amongst the $\frac{1}{N}{\rm tr} W^n$,
together with factorization (\ref{factor}) and  shape independence
(\ref{shp}) are solved by $\langle\frac{1}{N}{\rm tr} W^n\rangle =
\frac{1}{N}{\rm tr} M^n$, where $M$ is an $N\times N$ matrix
satisfying ${\rm tr} M^n = {\rm tr} M$.   One set of solutions are
the projection matrices which satisfy $M^n =M$.  However the only
solution of ${\rm tr} M^n = {\rm tr} M$ which  is smoothly related
to the weak coupling result, $\langle\frac{1}{N}{\rm tr}
W^n\rangle|_{g=0} = 1$, corresponds to $M=I$ such that
$\langle\frac{1}{N}{\rm tr} W^n\rangle =1$.

These results can be trivially extended to maximally supersymmetric
Yang-Mills in $3,5$ and $6$ dimensions,  by using a four supercharge
superspace with dimension $0,2$ and $3$ respectively. In these cases
the gauge coupling is dimensionful and the result (\ref{result}) can
be extracted from shape invariance either by shrinking the Wilson loop
in $3$ dimensions, or expanding it in $5$ and $6$ dimensions.  In
7-dimensional maximally supersymmetric Yang-Mills,  one encounters a
non-trivial generalized Konishi anomaly\footnote{This anomaly is
important for a Dijkgraff-Vafa conjecture proposed in
\cite{Dijkgraaf:2003xk} via the arguments of \cite{Cachazo:2002ry}.},
while in higher dimensions $d>7$, there is no four-supercharge
subalgebra of dimension $d-3$ so our formalism is not applicable.

\section{Chiral Wilson lines in ${\cal N}=2$ SYM}

A similar non-renormalization theorem applies to Wilson lines in four
dimensional ${\cal N}=2$ Yang-Mills theories,  and is obtained by
using ${\cal N}=2$, $d=3$ superspace.  In terms of this superspace,
the ${\cal N}=2$, $d=4$ vector multiplet is comprised of vector
superfields $V(x^2)$ and adjoint chiral superfields $\Phi(x^2)$, where
the superspace spans $x^0,x^1,x^3, \theta, \bar\theta$. However unlike
the usual ${\cal N}=1$, $d=4$ superspace formulation, $V$ contains
only the components $A_{0,1,3}$ of the gauge connections  while $A_2$
is contained in $\Phi$.  The two adjoint hermitian scalars $X,Y$ are
split between $V$ and $\Phi$.  The bottom component of $\Phi$ is $A_2
+ i X$. As discussed previously, hypermultiplets of ${\cal N}=2$,
$d=4$ are comprised of pairs ${\cal N}=2$, $d=3$ chiral superfields
$Q^1(x^2)$ and $Q_2(x^2)$.

If there are hypermultiplets in the fundamental representation, we
can define a chiral Wilson lines extended in the $x^2$ direction,
of the form
\begin{align}Q_1^m(0)\,{\cal P}\exp\left(i\int_{0}^{X^2}
dx^2 \Phi\right)Q_2^n(X^2)\,,
\end{align}
where $m$ and $n$ are flavor indices.  The bottom component belongs to
the ${\cal N}=2, d=3$ chiral ring.  The properties of the chiral ring
are determined by the superpotential,  which is
\begin{align}\label{superp}
{\rm W} = \int dx^2 \tilde Q_i (\frac{\partial}{\partial x^2}
- i\Phi) Q_i \, .
\end{align}
In the absence of a Konishi anomaly,
\begin{align}
&\frac{\partial}{\partial X^2}\langle\left.
Q_1^m(0)\, {\cal P} \exp\left(i\int_{0}^{X^2} dx^2 \Phi\right)
Q_2^n(X^2)\right|_{\theta=\bar\theta =0} \rangle  \nonumber \\
&=\langle\left. Q_1^m(0)\,{\cal P}\exp\left(i\int_{0}^{X^2} dx^2
\Phi\right)\frac{\delta W}{\delta
Q_1^n(X^2)}\right|_{\theta=\bar\theta =0} \rangle = 0 \, .
\end{align}
Expectation values of this particular class of Wilson line have no
dependence on the length of the Wilson line.  Since the scalar
components of $Q_i^m$ belong to ${\cal N}=2$, $d=4$ hypermultiplets,
these expectation values are equivalent to expectation values of
local operators parameterizing the Higgs branch.  These results
are readily generalized to eight supercharge Yang-Mills theories
in $1,2$ and $3$ dimensions.  In five dimensions one encounters a
Konishi anomaly,  while for dimension $d>5$ a four supercharge
$d-1$ dimensional subalgebra does not exist.

\end{document}